\documentclass[amssymb,aps,floatfix,floats,preprintnumbers,twocolumn,superscriptaddress,nofootinbib]{revtex4}
\usepackage{rcs}
\usepackage{color,graphicx}
\usepackage{dcolumn}
\usepackage{bm}
\usepackage[normalem]{ulem}
\usepackage{chngcntr}
\usepackage[thinspace,thinqspace]{SIunits}
\usepackage{natbib}
\usepackage{amsmath}

\newcommand{\muo}{$\mu_{\textrm{0}}$}

\newcommand{\CRA}{CeRh$_{2}$As$_{2}$}

\newcommand{\Tc}{$T_{\textrm{c}}$}
\newcommand{\To}{$T_{\textrm{0}}$}

\newcommand{\TN}{$T_{\textrm{N}}$}

\begin{document}
\preprint{APS/123-QED}
\title{Decoupling multi-phase superconductivity from normal state ordering in \CRA}

\author{K. Semeniuk}
\email[Corresponding author:~]{konstantin.semeniuk@cpfs.mpg.de}
\author{D. Hafner}
\affiliation{Max Planck Institute for Chemical Physics of Solids, D-01187 Dresden, Germany}
\author{P. Khanenko}
\author{T. L\"uhmann}
\author{J. Banda}
\author{J. F. Landaeta}
\affiliation{Max Planck Institute for Chemical Physics of Solids, D-01187 Dresden, Germany}
\author{C. Geibel}
\author{S. Khim}
\affiliation{Max Planck Institute for Chemical Physics of Solids, D-01187 Dresden, Germany}
\author{E. Hassinger}
\email[Corresponding author:~]{elena.hassinger@tu-dresden.de}
\affiliation{Technical University Dresden, Institute for Solid State and Materials Physics, 01062 Dresden, Germany}
\author{M. Brando}
\email[Corresponding author:~]{manuel.brando@cpfs.mpg.de}
\affiliation{Max Planck Institute for Chemical Physics of Solids, D-01187 Dresden, Germany}
\date{June 20, 2023}
\begin{abstract}
\CRA\ is a multi-phase superconductor with \Tc\ = 0.26\,K. The two superconducting (SC) phases, SC1 and SC2, observed for a magnetic field $H$ parallel to the $c$ axis of the tetragonal unit cell, have been interpreted as even- and odd-parity SC states, separated by a phase boundary at \muo$H^{*}=4$\,T. Such parity switching is possible due to a strong Rashba spin-orbit coupling at the Ce sites located in locally non-centrosymmetric environments of the globally centrosymmetric lattice. Existence of another ordered state (Phase I) below a temperature \To\ $\approx$ 0.4\,K suggests an alternative interpretation of the $H^{*}$ transition: It separates a mixed SC+I (SC1) and a pure SC (SC2) state. Here, we present a detailed study of higher quality single crystals of \CRA, showing much sharper signatures at \Tc\ = 0.31\,K and \To\, = 0.48\,K. We refine the $T$-$H$ phase diagram of \CRA\ and demonstrate that $T_{0}(H)$ and $T_{\textrm{c}}(H)$ lines meet at \muo$H\approx6$\,T, well above $H^{*}$, implying no influence of Phase I on the SC phase switching. A basic analysis with the Ginzburg-Landau theory indicates a weak competition between the two orders.
\end{abstract} 

\maketitle

\CRA\ is a newly discovered heavy-fermion superconductor with remarkable properties. At temperatures below \Tc\ = 0.26\,K, the material shows an exceptionally rare multi-phase superconductivity~\cite{khim2021}. Apart from UPt$_{3}$~\cite{fisher1989,bruls1990,adenwalla1990} and UTe$_{2}$~\cite{braithwaite2019,aoki2020,kinjo2022preprint,sakai2022preprint}, virtually all other unconventional superconductors host only one superconducting (SC) state. Above \Tc, the compound shows a non-Fermi-liquid temperature dependence of specific heat, $C/T \propto T^{-0.6}$, and electrical resistivity, $\rho(T) \propto \sqrt{T}$, suggesting proximity to a quantum critical point (QCP) of unknown nature~\cite{khim2021} or influence of two-channel Kondo physics~\cite{ludwig1991}. Below the temperature of $T_{0}\approx0.4$\,K, the material also hosts a peculiar state, Phase I, proposed to be a unique case of a quadrupole-density-wave (QDW) order~\cite{hafner2022}, contrasting all known multipolar orders observed in Ce-based systems which are typically of local origin~\cite{effantin1985,kitagawa1996}.  Moreover, very recent nuclear quadrupolar and magnetic resonance experiments report broadening of the As lines below \TN\ $<$ \Tc\ at one of the two As sites, indicating the presence of internal fields, likely due to an antiferromagnetic (AFM) order in one of the two SC phases~\cite{kibune2022,ogata2023}.

\begin{figure}[b]
	\begin{center}
		\includegraphics[width=\columnwidth]{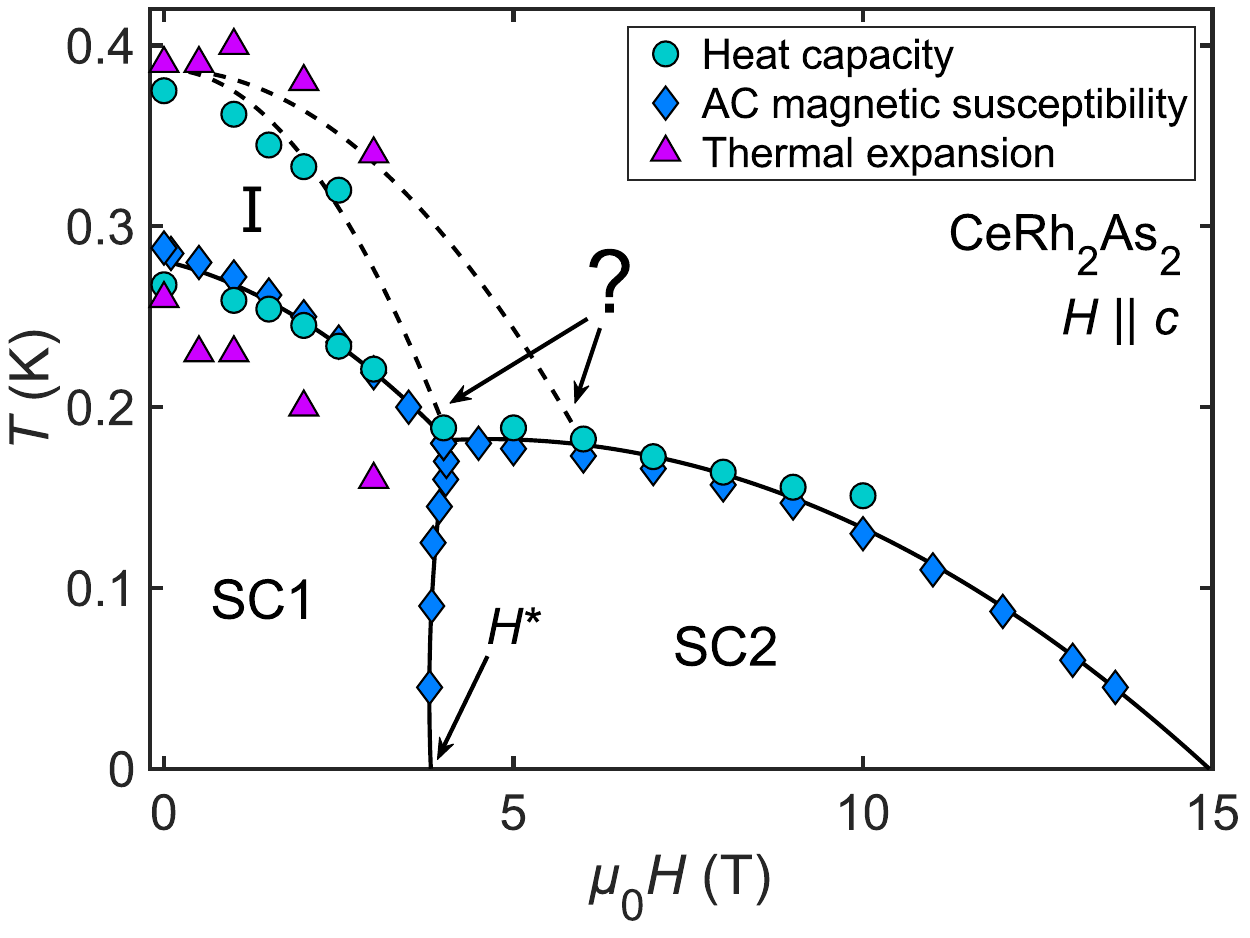}
		\caption{Temperature - $c$-axis magnetic field ($T$-$H$) phase diagram of \CRA\ according to magnetic and thermodynamic bulk probes. All measurements were conducted before this work, on the same batch of crystals (published in Refs.~\cite{khim2021,landaeta2022}, except the thermal expansion). The dashed lines highlight the ambiguity of the phase diagram: the phase-I transition line may meet the superconducting region either at the multicritical point at \muo$H^{*}=4$\,T, or at some higher field.}
		\label{fig1}
	\end{center}
\end{figure}

In fact, \CRA\ shows a single SC phase when magnetic field is applied along the basal plane of the tetragonal crystalline structure, but two SC phases, SC1 and SC2, for a field $H$ parallel to the $c$ axis (the relevant field direction for the rest of this paper). These phases are separated by a boundary at \muo$H^{*} = 4$\,T, as shown in Fig.~\ref{fig1}. The anisotropic field response of the superconductivity, as well as the presence of two SC phases in \CRA\ can be attributed to a strong Rashba spin-orbit coupling due to the locally non-centrosymmetric environments of Ce sites and quasi-two-dimensional character of the Fermi surface (FS)~\cite{khim2021,hafner2022,cavanagh2022}. The existing model is based on a scenario proposed ten years ago for layered structures comprising loosely coupled superconducting layers with alternating sign of the Rashba spin-orbit coupling. Such systems were predicted to exhibit a SC phase diagram very similar to that observed in \CRA~\cite{yoshida2012,sigrist2014,mockli2018,schertenleib2021,skurativska2021,nogaki2021,cavanagh2022,nogaki2022}, with the SC1-SC2 transition being a first order phase transition between even- and odd-parity SC order parameters~\cite{khim2021}. In the case of \CRA, such interpretation is also corroborated by the angle dependence of the SC upper critical field~\cite{landaeta2022}.

Besides the parity switching, there currently exist two alternative explanations for the multi-phase superconductivity in \CRA: 1) A field-induced magnetic transition within the SC state~\cite{machida2022}, in line with a recent NMR study~\cite{ogata2023}, and loosely reminiscent of the case of high-pressure CeSb$_{2}$~\cite{squire2023preprint}. 2) A transition between a mixed SC+I state (SC1) and a pure SC state (SC2). As illustrated in Fig.~\ref{fig1}, the $T_{0}(H)$ line curves towards lower temperature with increasing field and seems to vanish near the $T_{\textrm{c}}(H^{*})$ critical point~\cite{khim2021,landaeta2022,mishra2022}. If the second scenario is valid, all four phase boundaries should meet at a single multi-critical point. This can be verified by refining the phase diagram of \CRA. The topology of the phase diagram close to the $T_{\textrm{c}}(H^{*})$ point also has important thermodynamic implications. If the $T_{0}(H)$ and $T_{\textrm{c}}(H)$ lines meet at $H>H^{*}$, the SC1-SC2 boundary must be of the first order~\cite{yip1991}, and the two SC states must have distinct symmetries. These reasonings would hold regardless of the microscopic nature of Phase I.

In our experiments so far, insufficient sample homogeneity and broadening of features in magnetic field prevented us from reliably tracking Phase I boundary close to the multicritical point. Progress on this front therefore demands samples of higher quality.

Here, we report a detailed study of specific heat and electrical resistivity of higher quality single crystals of \CRA\ with \Tc\ = 0.31\,K and \To\, = 0.48\,K, which exhibit much clearer signatures of the phase transitions. We refine the $T$-$H$ phase diagram of \CRA\ and demonstrate that Phase I boundary meets the SC state at around 6\,T --- a significantly larger field than $H^{*}$ --- thus decoupling the SC1-SC2 switching and the presence of Phase I.  We do not observe any signatures of \To\ above 6\,T within the SC2 phase, yet sudden disappearance of Phase I at the SC2 phase boundary would go against thermodynamic principles. We also do not detect any signatures of magnetic transition within the SC states.

\begin{figure}[t]
	\begin{center}
		\includegraphics[width=\columnwidth]{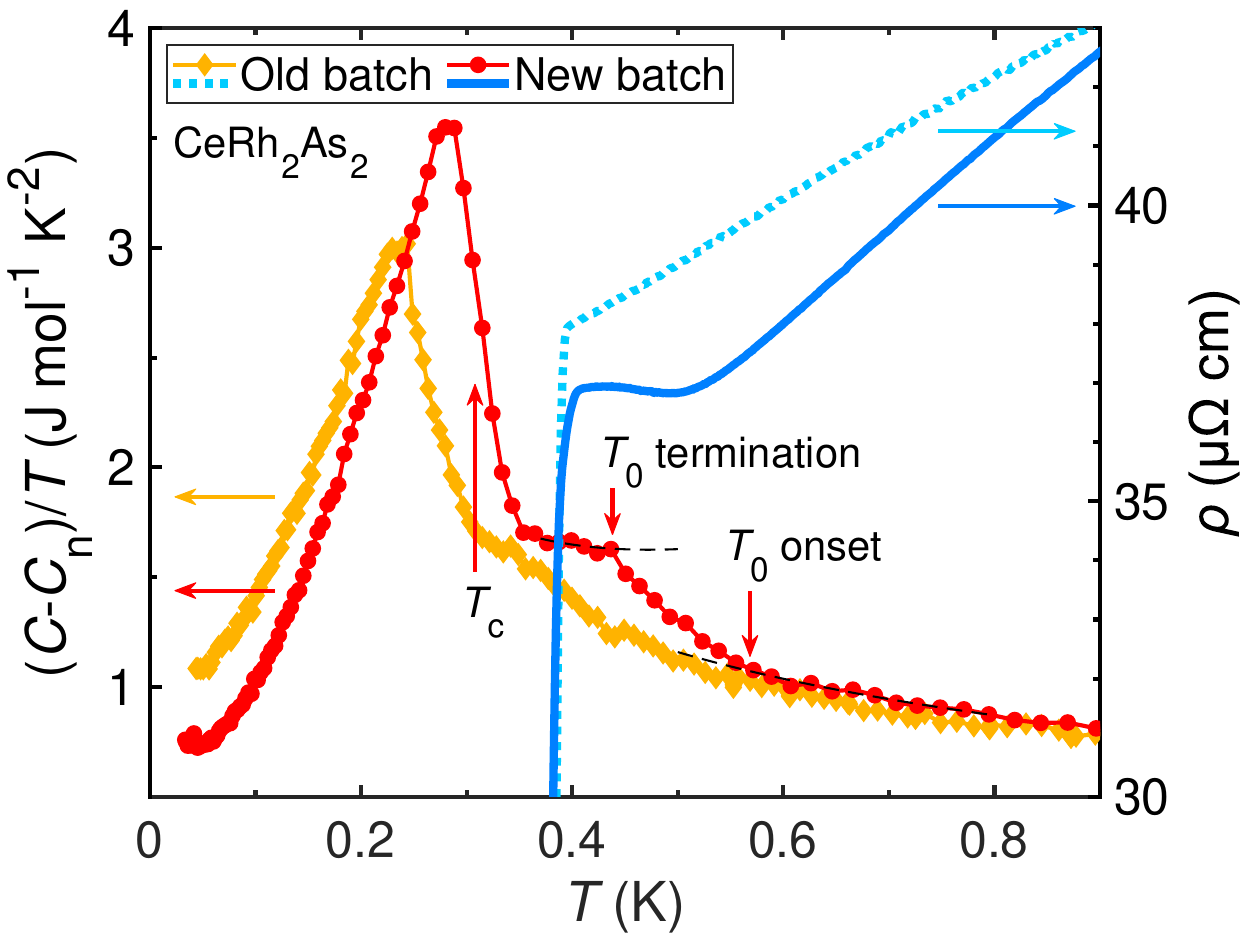}
		\caption{Specific heat $C(T)/T$ and resistivity $\rho(T)$ of \CRA\ as functions of temperature $T$ showing signatures of the superconductivity and Phase I (\Tc\ and \To, respectively) for two different generations of samples ("Old batch" denotes the samples used in Refs.~\cite{khim2021,hafner2022,landaeta2022}, while samples from the "New batch" were used in this work). The dashed black lines extrapolate sections of $C(T)/T$ for emphasising the phase-I transition. The nuclear specific heat contribution $C_{\mathrm{n}}(T)$ due to As atoms was subtracted (see Supplemental Material). Resistivity of the old batch sample is scaled by a factor of 0.6.}
		\label{fig2}
	\end{center}
\end{figure}

A significant improvement of the sample quality is illustrated in Fig.~\ref{fig2}, where thermodynamic and transport signatures of \To\ and \Tc\ are compared for the samples of \CRA\ used in Refs.~\cite{khim2021,hafner2022,landaeta2022,onishi2022} (old batch) and the new generation of samples studied in this work (new batch); see Supplemental Material (SM) for notes on the synthesis. One can identify two specific heat ($C(T)$) anomalies---the larger jump due to the SC transition and the smaller phase-I anomaly---both of which appear as typical second-order phase transitions, affected by broadening. Samples of the new batch show a higher bulk \Tc\ (0.31\,K against 0.26\,K) with a much sharper peak in specific heat, indicating a substantially improved homogeneity of crystals. The height of the jump at \Tc\ is $\Delta C/C|_{T_{\textrm{c}}} \approx 1.3$, whereas the value of 1 was reported for the old samples~\cite{khim2021}. The Sommerfeld coefficient $\gamma = C/T$ for $T \rightarrow 0$ decreases to 0.7\,J\,K$^{-2}$\,mol$^{-1}$ in the new sample, which is a 40\% reduction compared to the previous value of about 1.2\,J\,K$^{-2}$\,mol$^{-1}$~\cite{khim2021}, signifying a particularly strong sensitivity of superconductivity to disorder. The phase-I transition is likewise more pronounced, with its onset and termination clearly visible as changes in the slope of $C(T)/T$. The strong increase of \To\ from 0.4\,K to 0.48\,K with increasing sample quality is consistent with an itinerant nature of a sign changing order in $k$-space such as the proposed QDW.

Measurements of electrical resistivity ($\rho(T)$) reveal a visibly higher \Tc\ compared to the bulk value, reflecting the inhomogeneity of samples, also seen in other heavy-fermion systems~\cite{bachmann2019}. The discrepancy is reduced for the new batch, as the resistive \Tc\ of 0.38\,K is effectively the same as for the old batch~\cite{khim2021,hafner2022}. More importantly, the higher \To\ in the new batch makes the corresponding transport signature (resistivity upturn) much more apparent. Measurements of $\rho(T)$ can therefore be used for tracing the boundary of Phase I, which was rather challenging previously, as the transition at \To\ got rapidly obscured by the resistivity drop at \Tc\ upon applying field. The residual resistivity ratio (RRR) increases from $\sim$1.3 in the old batch to $\sim$2.8 in the new batch.

\begin{figure}[t]
	\begin{center}
		\includegraphics[width=\columnwidth]{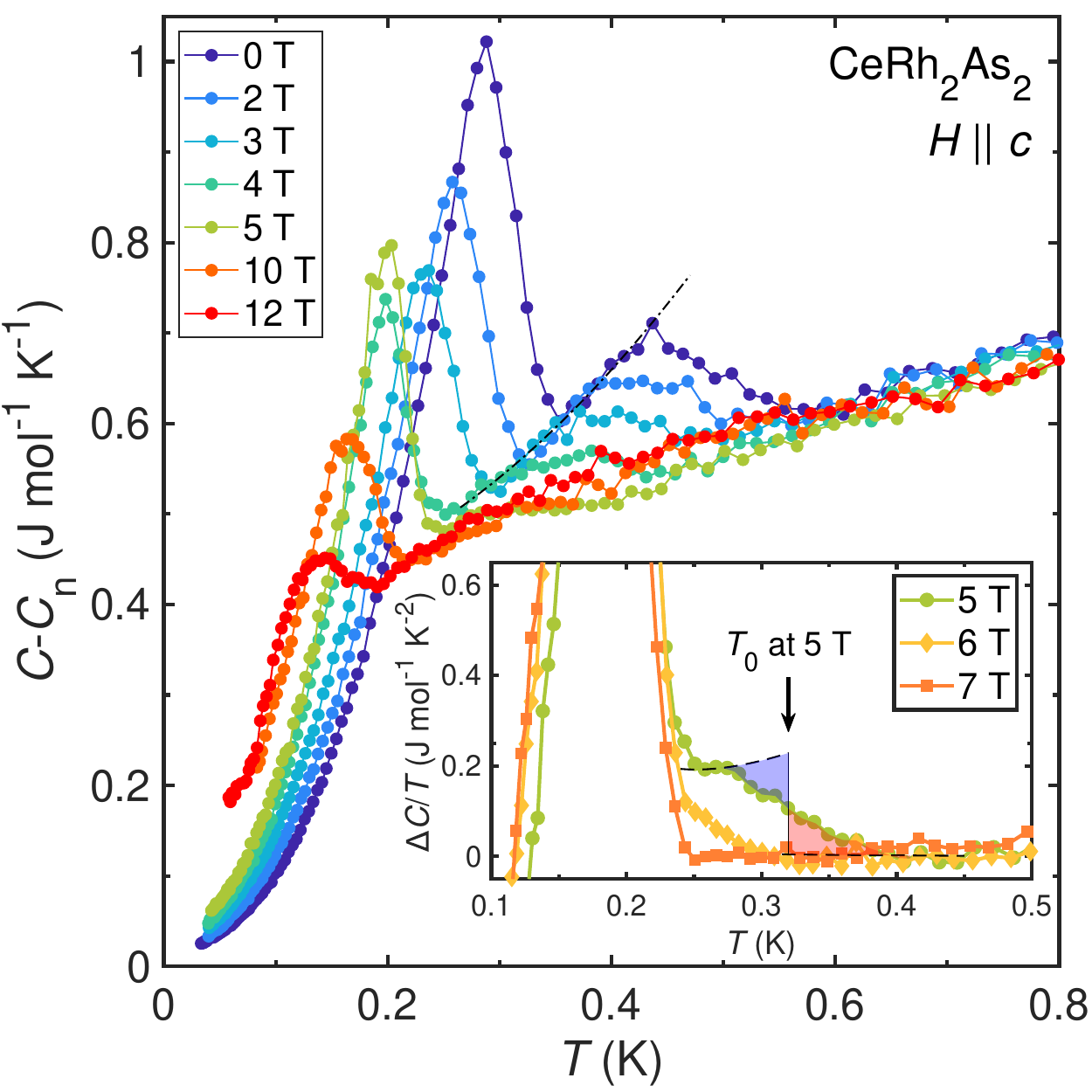}
		\caption{Electronic specific heat $C-C_{\mathrm{n}}$ of \CRA\ (new batch) against temperature $T$, measured at different $c$-axis magnetic fields. The dash-dot line marks the envelope curve of $C-C_{\mathrm{n}}$ between the onset of superconductivity and the termination of phase-I transition. Inset: disappearance of the phase-I transition signature in specific heat. The extrapolated unordered state specific heat was subtracted to emphasize the anomalies. For 5\,T, a plateau between 0.25\,K and 0.28\,K indicates that the Phase I transition is complete before the superconductivity sets in, and the critical temperature $T_{0}$ can be determined via the shown equal-entropy construction.}
		\label{fig3}
	\end{center}
\end{figure}

The low-temperature specific heat of \CRA\ at different fields is shown in Fig.~\ref{fig3} (see SM for the extended set of data). Besides the pronounced peaks at $T_{\textrm{c}}$ and $T_0$, there are no visible signatures of other phase transitions. Therefore, existence of the AFM order inside the SC phase~\cite{kibune2022} would imply that $T_{\textrm{c}}=T_{\textrm{N}}$. Both $T_{\textrm{c}}$ and $T_0$ anomalies shift to lower temperatures upon applying magnetic field, with the \To\ signature no longer apparent above 5\,T. We used the equal entropy construction for rigorously defining \Tc\ and \To\ (detailed in SM).

The specific heat jump height at \Tc\ exhibits a sharp increase as a function of field in a small interval around 4\,T because of the kink in the SC phase boundary line at the multicritical point at $H^{*}$ (as discussed in Ref.~\cite{khim2021} and demonstrated in SM). It is also instructive to look at the evolution of specific heat for 5\,T $\leq $ \muo$H \leq 7$\,T, shown in the inset of Fig.~\ref{fig3}. In this field range, \Tc\ is nearly constant. The phase-I transition is complete before the onset of superconductivity at 5\,T, is interrupted at 6\,T, and fully vanishes at 7\,T. At the same time, the peak value of specific heat at \Tc\ remains constant with respect to the unordered state ($T>T_{0}$). Disappearance of the \To\ signature between 5\,T and 6\,T is accompanied by a slight increase of specific heat below \Tc.

\begin{figure}[b]
	\begin{center}
		\includegraphics[width=\columnwidth]{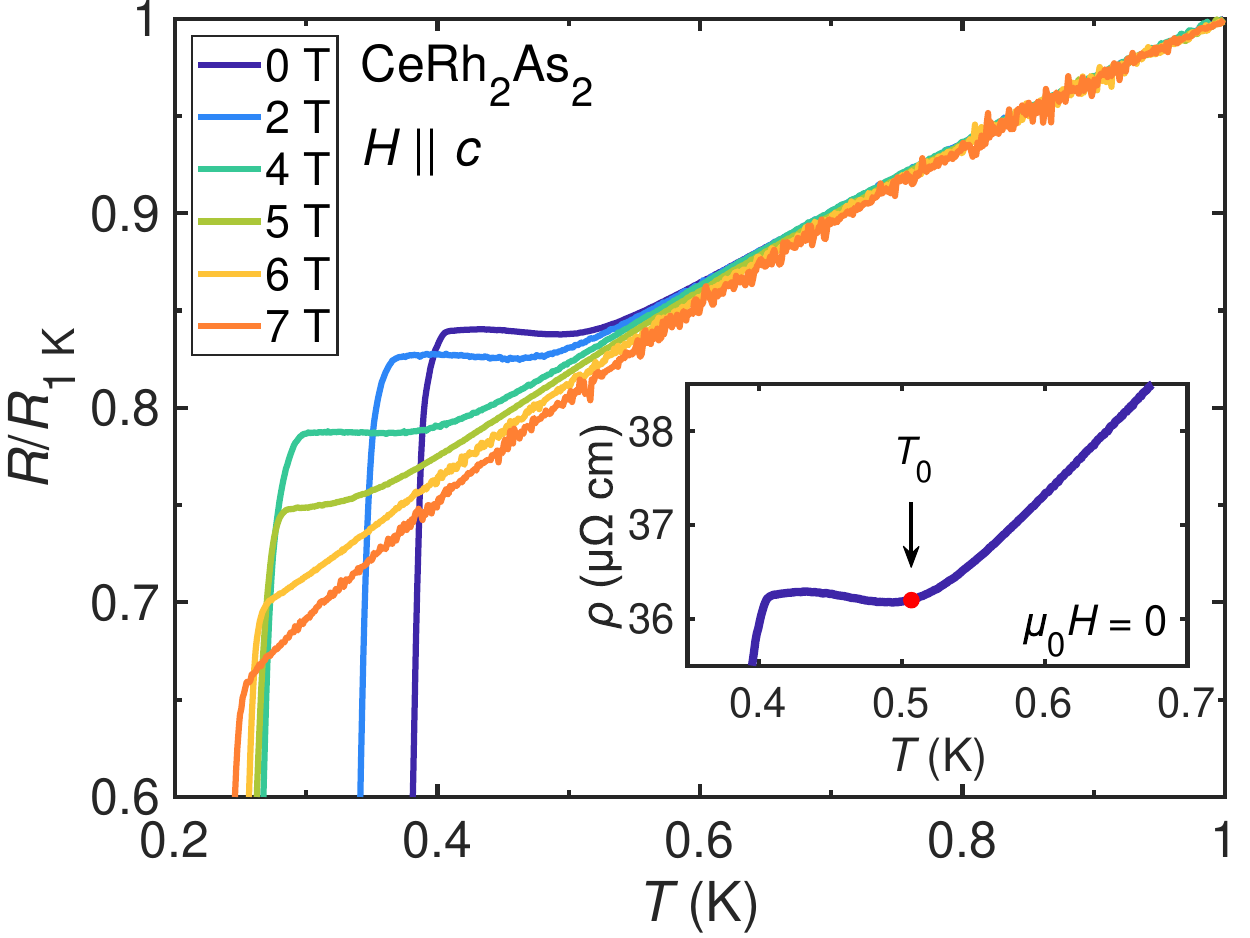}
		\caption{Electrical resistance $R$ against temperature $T$ for \CRA\ samples of the new batch, normalized to the resistance at 1\,K, for different $c$-axis magnetic fields. The inset displays zero-field resistivity $\rho(T)$ at the phase-I transition. The critical temperature \To\ was defined as the point of largest curvature (red dot).}
		\label{fig4}
	\end{center}
\end{figure}

We also investigated the behavior of the superconductivity and Phase I in field by measuring electrical resistivity for current parallel to the basal plane of the \CRA\ lattice. The resultant data for selected values of $H$ are shown in Fig.~\ref{fig4} (the extended set of data is available in SM). The phase-I transition is identifiable as a pronounced upturn in $\rho(T)$ below $\sim$ 0.6\,K. In zero field, upon subsequent cooling, the upturn is followed by a downturn and the eventual SC transition. We defined \To\ as the point of maximum curvature of $\rho(T)$ (red dot in the Fig.~\ref{fig4} inset; see SM for details on determining \To). The resultant boundary of Phase I decently reproduces that obtained from specific heat measurements, as will be shown next.

\begin{figure}[t]
	\begin{center}
		\includegraphics[width=\columnwidth]{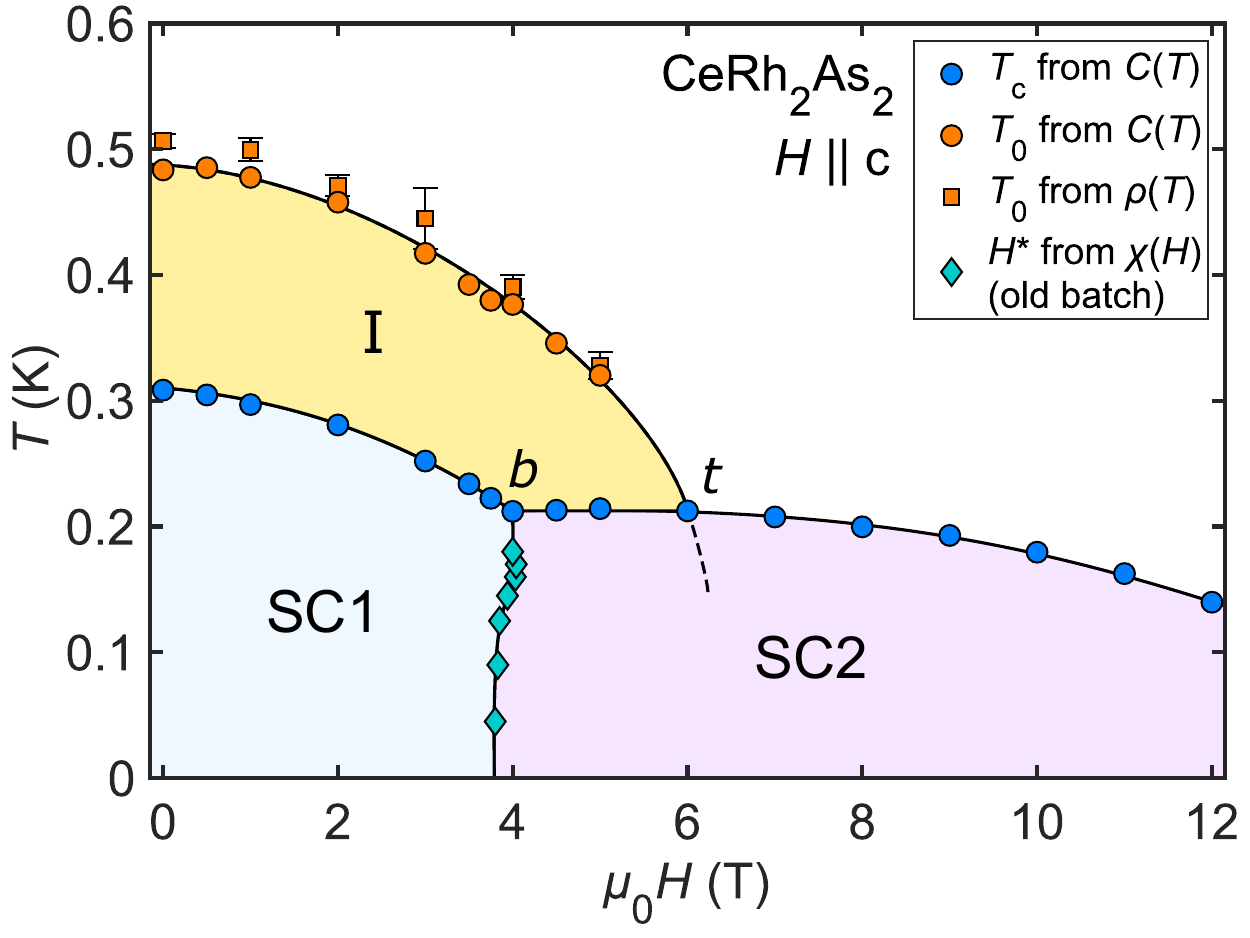}
		\caption{Temperature - $c$-axis magnetic field ($T$-$H$) phase diagram of \CRA\ depicting the two superconducting (SC) states SC1 and SC2 as well as Phase I. The SC and phase-I transition temperatures (\Tc\ and \To, respectively) come from measurements of specific heat $C(T)$ and electrical resistivity $\rho(T)$ conducted on samples from the new batch. The SC1-SC2 phase boundary, terminating in a bicritical point $b$, is plotted according to earlier AC magnetic susceptibility $\chi(H)$ data~\cite{khim2021}. The solid black lines are guides for the eye. The dashed line marks a segment of a so far undetected hypothetical phase boundary expected from thermodynamic considerations, while $t$ marks the corresponding tetracritical point.}
		\label{fig5}
	\end{center}
\end{figure}

We now discuss the $T$-$H$ phase diagram of \CRA\ shown in Fig.~\ref{fig5}. It summarizes the results of our specific heat and resistivity measurements conducted on samples from the new batch. The general shape of all phases---Phase I, SC1, and SC2---is consistent with the previously published data~\cite{khim2021,mishra2022}. While the transition temperatures are higher compared to the earlier generations of samples, the critical field \muo$H^{*}$ between SC1 and SC2 remains unchanged at 4\,T. The zero-temperature limit of the SC upper critical field extrapolates to roughly 15\,T and 18\,T for specific heat and resistivity respectively. We can clearly identify \To\ at fields as high as 5\,T, which was not possible in previous studies. The $T_{0}(H)$ line intercepts the SC2 phase boundary at approximately 6\,T. Consequently, we can definitively conclude that the boundary of Phase I does not meet the multicritical point at \muo$H^{*}=4$\,T and is not responsible for the associated phase transition.

According to thermodynamic considerations~\cite{yip1991}, it is not possible to have a multicritical point at some field $H_{c}$, such that two second order phase boundaries come out of it for $H<H_{c}$ and one phase boundary (first or second order) comes out for $H>H_{c}$ (holds true if the inequalities are swapped). Therefore, the point where the SC1 and SC2 phase boundaries meet is a bicritical one (labelled with the letter '\textit{b}' in Fig.~\ref{fig5}), and the SC1-SC2 transition must be of the first order. Applied to the crossing of the $T_{\textrm{c}}(H)$ and $T_{0}(H)$ lines at 6\,T, this constraint requires that the boundary of Phase I continues inside the SC2 state, resulting in a tetracritical point (indicated by the letter '\textit{t}' in Fig.~\ref{fig5}). That said, we do not observe any signature of Phase I within the SC2 phase (cf. 7\,T curve in the inset of Fig.~\ref{fig3}). Given the steep slope of the $T_{0}(H)$ line near 6\,T, it is quite possible that the transition evaded detection due to being too broad or the field sampling period of 1\,T being too large.

We analyzed the experimental phase diagram using the Ginzburg-Landau (GL) theory of coupled order parameters~\cite{imry1975}, as done, e.g., for iron pnictide superconductors in Ref.~\cite{fernandes2010}. The coupling term in the free energy between the SC2 and phase-I order parameters, $\Delta$ and $\textbf{Q}$ respectively, is given by $\lambda \Delta^{2}\mathbf{Q}^{2}$, where $\lambda$ indicates the strength of the coupling as well as its nature, $\lambda > 0$ for competing and $\lambda < 0$ for supporting coupling. We reproduced our experimental phase diagram by using the values of the transition temperatures, the slopes of the phase boundary lines and the jumps of the specific heat coefficients at the transition temperatures, $\Delta C(T_{0})/T_{0}$ and $\Delta C(T_{c})/T_{c}$, which are directly related to the condensation energies and to the changes in the slopes of the $T_{0}(H)$ and $T_{\textrm{c}}(H)$ lines at \textit{t}. A detailed analysis is provided in SM; here we summarize the main findings: i) The absence of a pronounced kink in the $T_{\textrm{c}}(H)$ line at \textit{t} implies that the coupling is not strong. ii) The supporting coupling is unlikely, since it flattens the $T_{0}(H)$ line below \textit{t} (Fig.~S9d), and we should have then observed a double feature in $C(T)/T$ at 7\,T. iii) In case of a weak competition, the slope change across \textit{t} would be about two orders of magnitude larger for the $T_{0}(H)$ line than for the $T_{\textrm{c}}(H)$ line (Fig.~S9a). This is due to the fact that the changes of the slopes at \textit{t} are related to $\Delta C(T_{0})/T_{0}$ and $\Delta C(T_{c})/T_{c}$ evaluated at \textit{t}, and in our case $\Delta C(T_{0})/T_{0} \approx \frac{1}{10}\Delta C(T_{c})/T_{c}$. For example, for $\lambda$ equal to 20\% of the value necessary to induce a first order transition, the critical field of Phase I at $T=0$ would go down by about 0.4\,T compared to the case of no coupling, resulting in no phase-I-related feature observable in specific heat at 7\,T. At the same time, $T_{\textrm{c}}$ should slightly increase between the \textit{b} and \textit{t} points. Below the line joining \textit{b} and \textit{t}, the two states would then homogeneously coexist~\cite{fernandes2010}. We indeed observe a slight change in the slope of $T_{\textrm{c}}$ at \textit{t} (see Fig.~S10), consistent with the weak coupling scenario, but extraction of quantitative information is hindered by the insufficient resolution of our data.

Whereas in iron pnictides both superconductivity and magnetism originate from the same FS sheet, the situation in \CRA\ is more complex due to the presence of two relevant sheets. The most reliable renormalized band structure calculations so far~\cite{hafner2022} predict a quasi-three-dimensional strongly-corrugated cylinder (3D-sheet) along the $\mathit{\Gamma}$-$Z$ direction of the Brillouin zone (BZ) and quasi-two-dimensional corrugated cylinders (2D-sheet) along axes parallel to the $A$-$M$ direction. Within the parity change picture of the SC1-SC2 transition, the SC pairing is mainly driven by the electrons of the 2D-sheet, located at the edges of the BZ, where the spin-orbit coupling dominates over the interlayer hopping~\cite{cavanagh2022}. A recent study~\cite{mishra2022} demonstrated that the upturn of $\rho(T)$ at \To\ is stronger when the current flows along the $c$ axis as opposed to the $ab$ plane. Within the QDW scenario, the associated FS nesting vector must therefore have a sizeable out-of-plane component, and the states involved in the QDW instability are expected to belong to the 3D-sheet of the FS. The attribution of the two orders to the different FS sheets disfavours a strong coupling between the QDW and superconductivity, and one would then expect the $T_{0}(H)$ line to continue into the SC2 phase practically undisturbed.

To conclude, using new higher quality crystals we refined the $T$-$H$ phase diagram of \CRA\ for field parallel to the $c$ axis, and showed that the Phase I boundary unambiguously does not meet the bicritical point between the two SC phases and is therefore not responsible for the SC1-SC2 transition. The $T_{0}(H)$ line intercepts the SC2 phase boundary at a tetracritical point near 6\,T. Our analysis of the phase boundaries near the tetracritical point suggests a weak competing interaction between the SC and phase-I order parameters. These conclusions leave us with two viable explanations for the origin of the SC1-SC2 transition: the even-to-odd parity switching~\cite{khim2021,landaeta2022} or a change in magnetic ordering~\cite{machida2022}. Our results prompt a further study of the 6\,T tetracritical point, from which an additional phase boundary is expected to emerge.
\begin{acknowledgments}
We are indebted to D. Agterberg, D. Aoki, S. Kitagawa, G. Knebel, K. Ishida, A. P. Mackenzie, A. Rost, O. Stockert, Y. Yanase and G. Zwicknagl for useful discussions. This work is also supported by the joint Agence National de la Recherche and DFG program Fermi-NESt through Grants No. GE602/4-1 (C. G. and E. H.). Additionally, E. H. acknowledges funding by the DFG through CRC1143 (project number 247310070) and the W\"urzburg-Dresden Cluster of Excellence on Complexity and Topology in Quantum Matter — ct.qmat (EXC 2147, project ID 390858490).
\end{acknowledgments}
\vspace{20pt}

\begin{center}
\textbf{\large Supplemental Material}
\end{center}

%%%%%%%%%% Merge with supplemental materials %%%%%%%%%%
%%%%%%%%%% Prefix a "S" to all equations, figures, tables and reset the counter %%%%%%%%%%
\setcounter{equation}{0}
\setcounter{figure}{0}
\setcounter{table}{0}
\makeatletter
\renewcommand{\theequation}{S\arabic{equation}}
\renewcommand{\thefigure}{S\arabic{figure}}

\section{C\lowercase{e}R\lowercase{h}\textsubscript{2}A\lowercase{s}\textsubscript{2} crystal growth}
We have partially modified the synthesis condition described in the previous work \cite{khim2021}. Elemental metals with the ratio of Ce:Rh:As:Bi = 1:2:2:30 were placed in an alumina crucible which was sealed in a Ta tube filled with 1~bar Ar. After being welded in an arc melter, the Ta tube wrapped with Zr foil was located in an Ar-filled vertical tube furnace. The furnace was heated to 1280\,\degree C for three days and slowly cooled down to 750\,\degree C at a rate of 2\,\degree C/h. Grown crystals were obtained after removing the Bi flux using diluted nitric acid. We found that the increased temperature and Ar partial pressure yielded quality-improved samples.

\section{Subtraction of nuclear contribution to heat capacity}
Here we evaluate the Schottky nuclear contribution to the specific heat of \CRA. Since a Ce atom has no nuclear moment and the contribution from the Rh nuclear moment is well below 1\,mK~\cite{pobell2007,steppke2010}, the only relevant contribution to the specific heat is from the $^{75}$As nuclear moment $I = 3/2$ (with 100\% abundance). \CRA\ has a tetragonal structure with two sites for the As atoms, As(1) and As(2). Similar analysis was done by Hagino \textit{et al.} for CuV$_{2}$S$_{4}$ (Cu nuclear spin $I = 3/2$)~\cite{hagino1994}.

The nuclear-spin Hamiltonian of an atomic nucleus with a spin quantum number $I$ and a magnetic moment $\mu_{I}=g_{N}\mu_{N}I$ with $\mu_{N}=5.05\times 10^{-27}$~J/T (nuclear magneton) and $g_{N}$ the nuclear $g$-factor, consists of the sum of quadrupolar and Zeeman terms:
\begin{equation}\label{eq1}
	\begin{split}
H & = \frac{e^{2}qQ}{4I(2I-1)}\left[3m_{I}^{2}-I(I+1)+\frac{1}{2}\eta(I_{+}^{2}+I_{-}^{2})\right]\\ 
  & - \gamma_{N}\hbar I \cdot B
	\end{split}
\end{equation}
where $\gamma_{N} = g_{N}\mu_{N}/\hbar$ is the gyromagnetic ratio and $m_{I}$ is the azimuthal quantum number with allowed values $I$, $I-1$, \ldots, $-I$. The equally spaced energy levels in an effective field $B_{eff}$ are:
\begin{equation}\label{eg2}
\varepsilon_{m}=-\mu_{I} \cdot B_{eff} = -g_{N}\mu_{N}m_{I}B_{eff}
\end{equation}
so the energy difference between the $2I+1$ levels is:
\begin{equation}\label{eq3}
\Delta\varepsilon=-g_{N}\mu_{N}B_{eff}.
\end{equation}
This energy gap is generally very small and the Schottky peaks usually occur at temperatures of the order of mK; consequently, in the temperature range of our measurements only the high-temperature part of the peak can be observed and the analysis can be made using only the proportional factor
$\alpha=C_{N}T^{2}$ which can be written as:
\begin{equation}\label{eq4}
\alpha=\frac{R}{3}I(I+1)\left(\frac{\hbar\gamma_{N}}{k_{B}}\right)^{2}B_{eff}^{2}
\end{equation}
where $\gamma_{N}=\omega_{N}/B_{eff}$ corresponds to an NMR-frequency $\omega_{N}$ which gives an energy splitting of $\hbar\omega_{N}$ in an effective magnetic field $B_{eff}$. It can be calculated/measured for every isotope and is of the order of $10^{6}$~Hz/T.

If the nucleus has a quadrupole moment $Q$ ($I \geq 3/2$) and is situated in a non-spherical or non-cubic electronic environment, its interaction with the field gradient $V_{zz} = eq$ with an anisotropy parameter $\eta$ (with axially symmetric field gradient, $\eta = 0$) produced by neighboring atoms will cause small splittings of the energy levels:
\begin{equation}\label{eq5}
\Delta\varepsilon=\varepsilon(m_{I})-\varepsilon(m_{I}')= \frac{3e^{2}qQ}{4I(2I-1)}|m_{I}^{2}-m_{I}'^{2}|
\end{equation}
and the proportional factor in the Schottky specific heat can be given by
\begin{equation}\label{eq6}
\alpha = \frac{R}{80}\left(\frac{e^{2}qQ}{k_{B}}\right)^{2}\frac{(I+1)(2I+3)}{I(2I-1)}.
\end{equation}

\begin{table}[t]
	\centering
	\caption{The $^{75}$As atom.}
	\begin{tabular}{ll}
		\hline\hline
		Parameters &  Constants\\
		$I = 3/2$ & $k_{B} = 1.38 \times 10^{-23}$\,J/K \\
		$Q = 0.3 \times 10^{-28}$\,m$^{2}$ & $e = 1.6 \times 10^{-19}$\,K \\
		$\gamma = 7.294$\,MHz/T & $h = 6.63 \times 10^{-34}$\,Js\\
		Abundance = 100\% & $R = 8.31$\,J/Kmol
	\end{tabular}
	\label{tab1}
\end{table}

\begin{figure}[t]
	\centering
	\includegraphics[width=0.8\columnwidth]{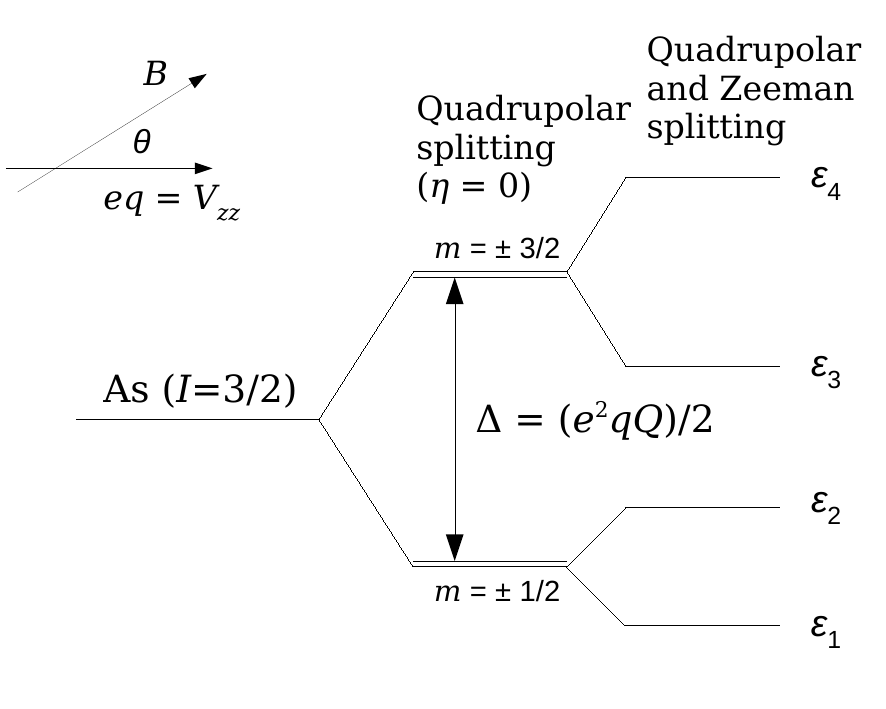}
	\caption{\label{fig:SI_level_scheme} The nuclear quadrupole splitting and the Zeeman splitting for the $^{75}$As nuclear spin $I$ in the absence and the presence of magnetic field $B$. The field gradient $eq = V_{zz}$ is axially symmetric ($\eta = 0)$.}
\end{figure}

In \CRA\ the only relevant contribution to the specific heat is from the $^{75}$As nuclear moment $I = 3/2$ (with 100\% abundance) and its level scheme is shown in Fig.~\ref{fig:SI_level_scheme}. All relevant parameters are listed in Tab.~\ref{tab1}.

In zero magnetic field, the energy difference given by Eq.~\ref{eq5} is:
\begin{equation}\label{eq7}
\Delta = \frac{1}{2}e^{2}qQ
\end{equation}
which corresponds to an NQR frequency $\nu_{Q} = \Delta/h$.

The energy levels $\epsilon_{i}$ can be calculated:
\begin{align*}
\epsilon_{1} & = -\frac{\Delta}{2} - \frac{1}{2}\gamma_{N}\sqrt{4-3\cos^{2}(\theta)}\\
\epsilon_{2} & = -\frac{\Delta}{2} + \frac{1}{2}\gamma_{N}\sqrt{4-3\cos^{2}(\theta)}\\
\epsilon_{3} & = +\frac{\Delta}{2} - \frac{3}{2}\gamma_{N}\cos(\theta)\\
\epsilon_{4} & = +\frac{\Delta}{2} + \frac{3}{2}\gamma_{N}\cos(\theta)		
\end{align*}
and the high-term part of the Schottky molar heat capacity that we can observe in the experiment is:
\begin{equation}\label{eq8}
\alpha = C_{\textrm{n}}T^{2} = \frac{R}{4}\left(\frac{\Delta}{k_{B}}\right)^{2} = R \left( \frac{e^{2}qQ}{4k_{B}} \right)^{2}.
\end{equation}
In magnetic field $B(T)$ with $\theta = 0$ it becomes:
\begin{equation}\label{eq9}
\alpha(B) = C_{\textrm{n}}(B)T^{2} = \frac{5R}{4}\left(\frac{\gamma_{N}\hbar B}{k_{B}}\right)^{2}.
\end{equation}

\begin{figure}[t]
	\centering
	\includegraphics[width=\columnwidth]{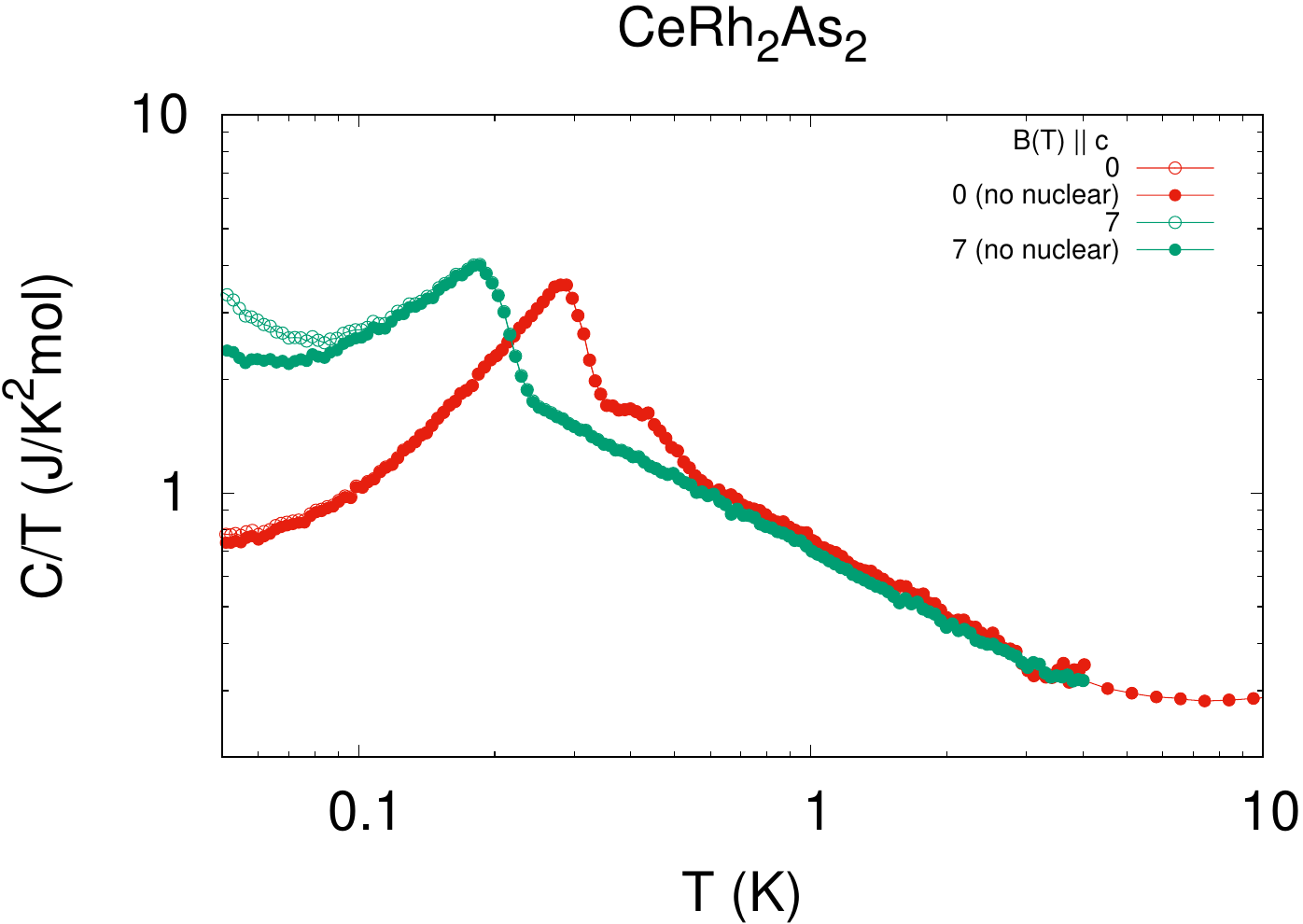}
	\caption{\label{fig:SI_C_nuclear} Specific heat plotted for $B = 0$ and 7\,T with and without nuclear contribution.}
\end{figure}

We can now use the measured NQR frequencies from Kibune \textit{et al.}~\cite{kibune2022} to calculate the energy splittings:
\begin{align*}
&As(1):\\
&	\nu_{Q} = \Delta/h = 31.1\,\mathrm{MHz}\\
&	\Delta = 2.06 \times 10^{-26}\,\mathrm{J} = 1.49\,\mathrm{mK}\\
&As(2):\\
&	\nu_{Q} = \Delta/h = 10.75\,\mathrm{MHz}\\
&	\Delta = 0.71 \times 10^{-26}\,\mathrm{J} = 0.51\,\mathrm{mK}.
\end{align*}
Considering that the values for the energy splittings are very low (about 1\,mK), a very small contribution to the specific heat should be observed at temperatures between 50\,mK and 100\,mK in zero field, as shown in Fig.~\ref{fig:SI_C_nuclear}.

From these data we calculate the $\alpha$ coefficient using Eq.~\ref{eq8}:
\begin{align*}
	As(1)&: \alpha = 4.63 \times 10^{-6}\,\mathrm{JK/mol}\\
	As(2)&: \alpha = 0.55 \times 10^{-6}\,\mathrm{JK/mol}\\
	\mathrm{Total{-}value}&: \alpha = 5.18 \times 10^{-6}\,\mathrm{JK/mol}\\
\end{align*}

In magnetic field we can use the NMR frequency $\gamma_{N} = 2\pi\gamma = 45.8$\,MHz/T, and with Eq.~\ref{eq9} we obtain:
\begin{align*}
\mathrm{Total{-}value} &: \alpha(B) = 2.54 \times 10^{-6} \times B^{2}\,\mathrm{JK/molT^{2}}.
\end{align*}
\newpage
\section{Extended heat capacity data set}

\begin{figure}[h]
\includegraphics[width=\linewidth]{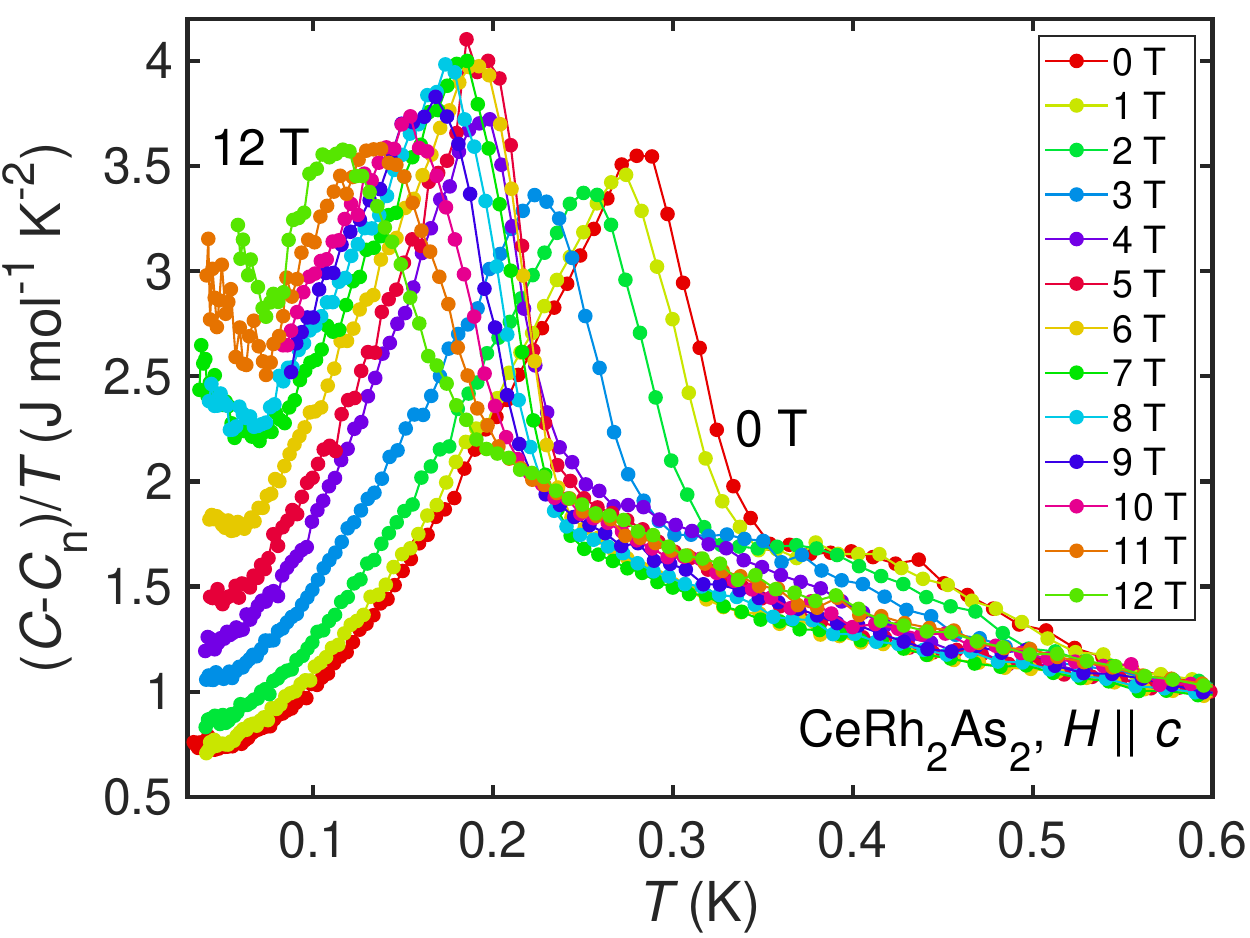}
\caption{\label{fig:SI_C_full} Temperature dependence of electronic specific heat $(C - C_{\mathrm{n}})/T$ of \CRA\ at different magnetic fields applied along the $c$-axis.}
\end{figure}

In Fig.~\ref{fig:SI_C_full} we show the raw data of specific heat of \CRA\ down to 40\,mK, for $c$-axis magnetic fields of up to 12\,T, with setps of 1\,T.

\section{Determination of $T_{\mathrm{c}}$ and $T_{0}$ using the equal entropy construction}

\begin{figure}[t!]
\includegraphics[width=\linewidth]{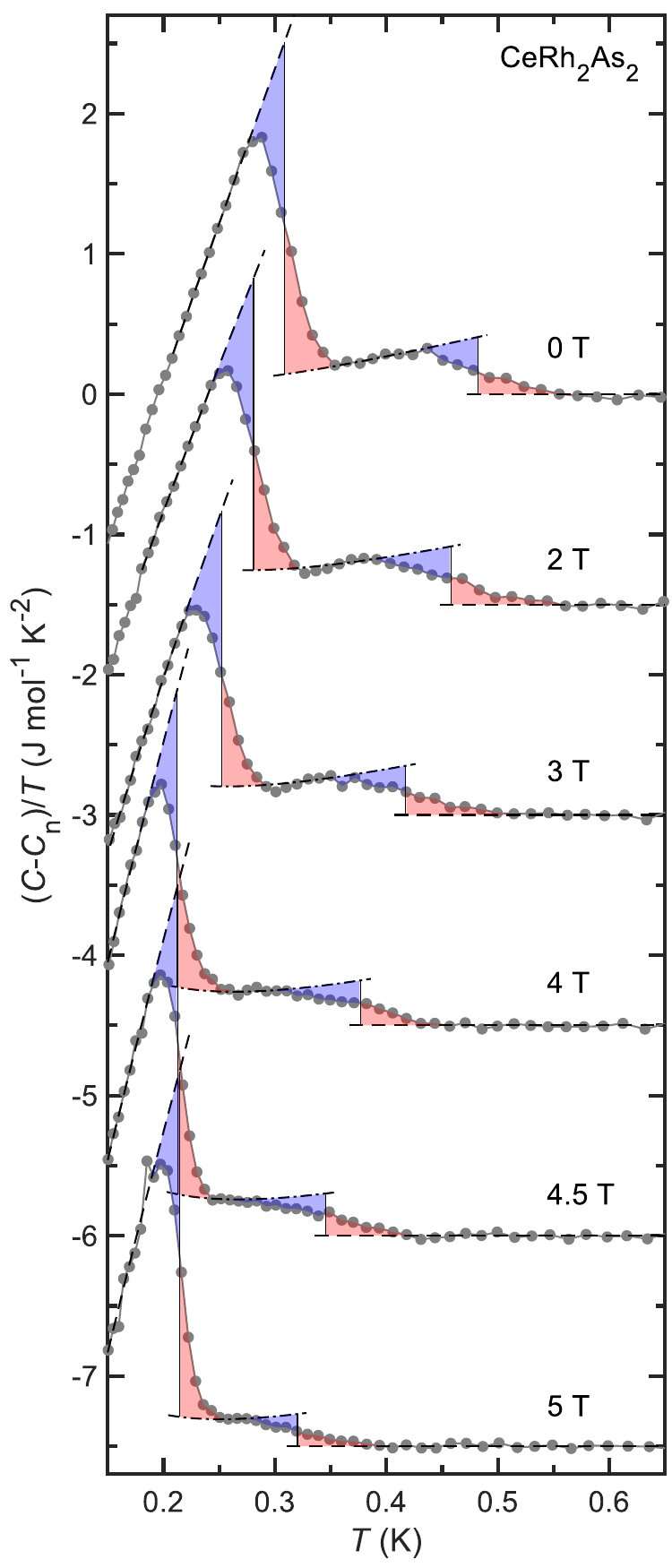}
\caption{\label{fig:SI_equal_entropy} Determination of \Tc\ and \To\ in \CRA\ from specific heat at different magnetic fields via the equal entropy method. The power-law extrapolated unordered state heat capacity was subtracted. The sets of data for different fields are offset from each other by 1.5\,J\,K$^{-2}$\,mol$^{-1}$. For each anomaly, the adjacent pink and purple shaded areas are equal.}
\end{figure}

When processing the specific heat data, we determined the values of \Tc\ and \To\ via the equal entropy construction. For a given second order transition, the critical temperature was defined as the temperature at which the entropy change for a hypothetical zero-width transition was the same as the observed entropy change for the real impurity-broadened transition. As visualized in Fig.~\ref{fig:SI_equal_entropy}, this means finding such values of \Tc\ or \To\ that equalize the integrals corresponding to the pink and purple shaded areas (each transition has to be treated separately). To definine the "ideal" specific heat (dashed and dash-dot lines in Fig.~\ref{fig:SI_equal_entropy}), one has to extrapolate parts of the measured data. In the case of Phase I, we note that between the onset of \Tc\ and the termination of \To, the measured specific heat effectively falls on the same curve (within a $<$0.02\,J\,K$^{-2}$\,mol$^{-1}$ constant offset) at all relevant fields up to and including 5\,T (dash-dot lines in Fig.~3 and Fig.~\ref{fig:SI_equal_entropy}). We approximated this curve by fitting a second order polynomial to the relevant set of $(T,(C-C_{\mathrm{n}}))$ points.

\section{Field dependence of the specific heat jump at the superconducting transition}

\begin{figure}[h]
\includegraphics[width=\linewidth]{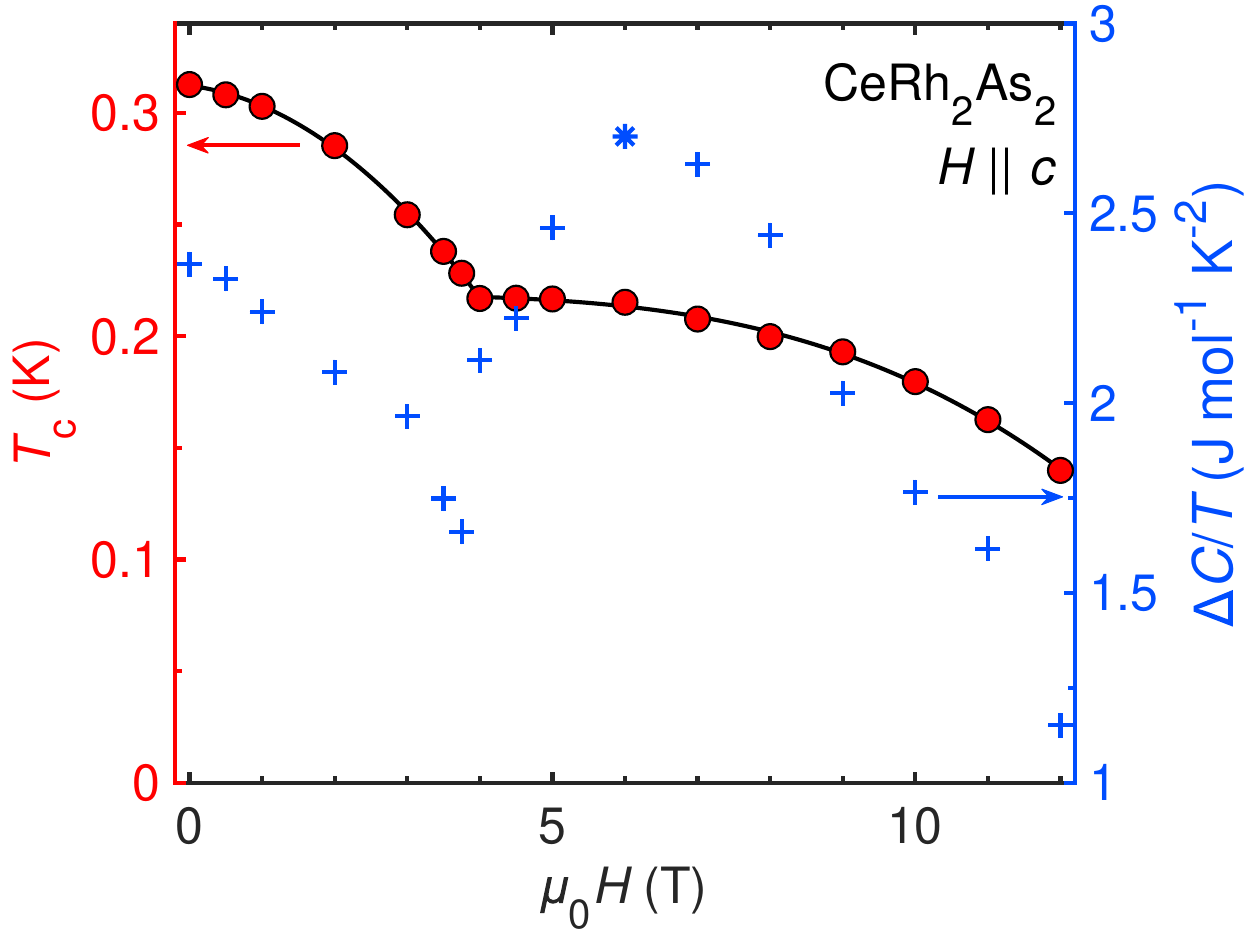}
\caption{\label{fig:SI_C_jump} Left axis: $c$-axis magnetic field dependence of \Tc, according to the heat capacity data. Right axis: the ideal specific heat ($C/T$) jump at \Tc.}
\end{figure}

Applying the equal entropy construction (described above) allowed us to determine the ideal specific heat jump heights $\Delta C/T$ at \Tc\ for different $c$-axis magnetc fields. In Fig.~\ref{fig:SI_C_jump} we plot the field dependence of the jump in $C/T$ at \Tc. The jump height increases suddenly as the field crosses the \muo$H^{*}=4$\,T value. This change is linked to the existence of the first order SC1-SC2 phase boundary at 4\,T~\cite{yip1991}. At 6\,T, the two transitions overlap, hindering the equal entropy analysis. For this particular point, we provide the $C/T$ jump height with respect to the extrapolated unordered state (the state above the onset of the phase-I transition). Since we observe the onset of the phase-I transition at 6\,T, the plotted value (* marker) overestimates the actual ideal $C/T$ jump height at the SC transition.

\section{Extended resistivity data set}

\begin{figure}[t]
\includegraphics[width=\linewidth]{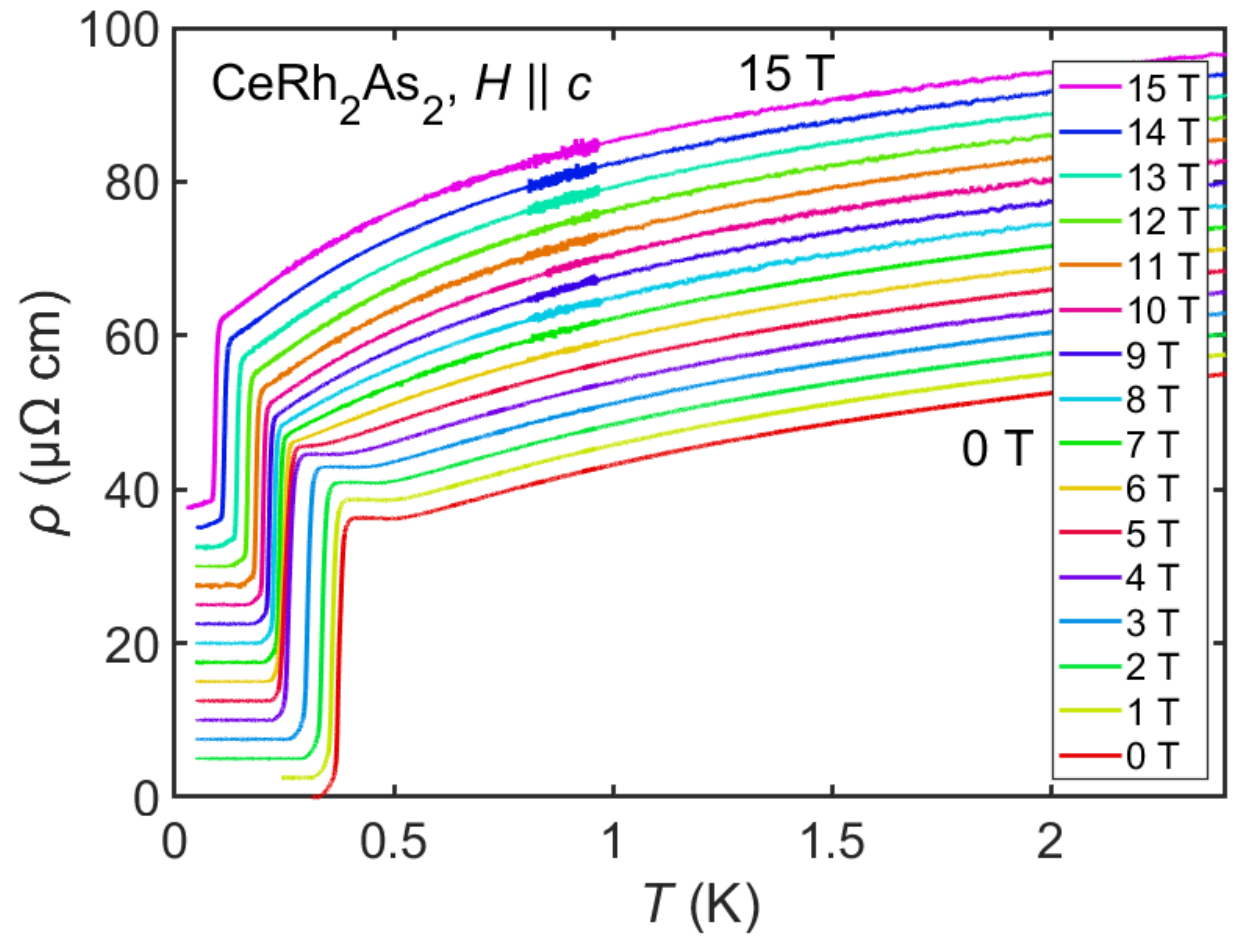}
\caption{\label{fig:SI_rho_full} Resistivity ($\rho$) of \CRA\ as a function of temperature ($T$) at different $c$-axis magnetic fields. The curves are offset from each other by 2.5\,$\micro\Omega$\,cm for improved clarity.}
\end{figure}

In Fig.~\ref{fig:SI_rho_full} we show the raw data of electrical resistivity of \CRA\ between 50\,mK and 2\,K for $c$-axis magnetic fields of up to 15\,T, with steps of 1\,T.

\section{Determination of \To\ from resistivity}

Resistivity upturn is a common signature of density-wave type transitions (e.g. Ref.~\cite{gruner2017}). When it occurs at temperatures well above 1\,K, the transition generally appears as a kink of negligible width compared to the temperature interval of interest. In our case, the transition takes place over significant portions of temperature sweeps, and deciding which particular temperature value to take as \To\ becomes non-trivial.

\begin{figure}[t]
\includegraphics[width=\linewidth]{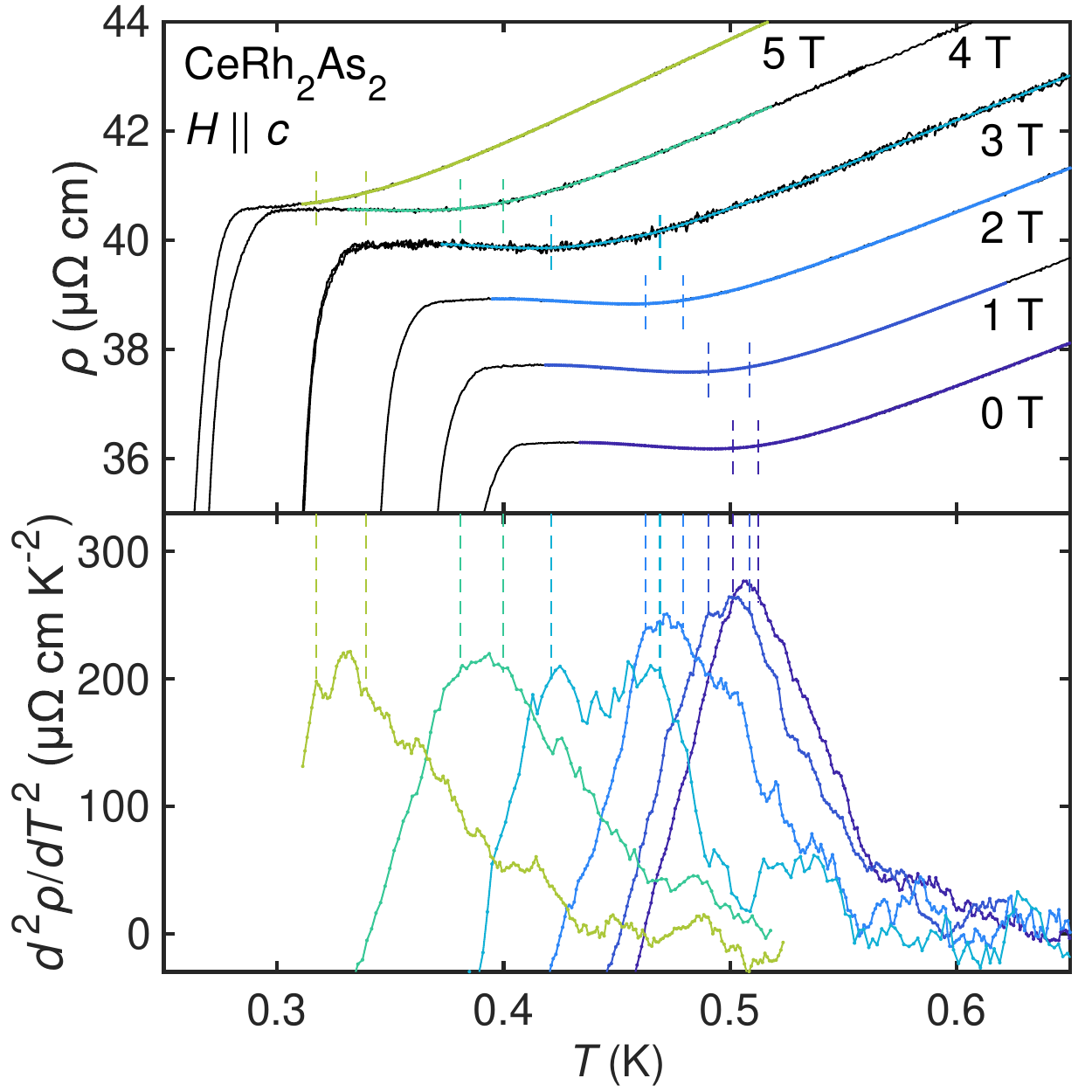}
\caption{\label{fig:SI_T0_resistivity_definition} Determination of the phase-I transition temperature \To\ according to measurements of resistivity $\rho(T)$. Top: raw resistivity (black curves) with smoothed data plotted on top for different $c$-axis magnetic fields. The curves are offset by 1.5\,$\micro\Omega$\,cm. Bottom: second derivatives of resistivity with respect to temperature. The vertical dashed lines mark the upper and lower bounds for locating the second derivative maxima. The smoothing and differentiation procedures are explained in the text.}
\end{figure}

We interpret the resistivity upturn as a partial gapping of the Fermi surface. It is therefore sensible to define the midpoint of the phase-I transition as the point where the curvature of $\rho(T)$, i.e. its second derivative, is the largest. This value of \To\ can be vaguely interpreted as the temperature at which the rate of charge carrier depletion is the highest. Numerical differentiation of $\rho(T)$ using the finite difference method produces exceesively noisy data. Instead, we determined the second derivative using the Savitzky-Golay smoothing procedure (shown in Fig.~\ref{fig:SI_T0_resistivity_definition}). Before processing a sweep, we first truncated it just above the onset of superconductivity, to ensure that we considered only the normal state resistivity. We then binned the $\rho(T)$ data into intervals of 1\,mK (0\,T, 1\,T, 2\,T, 5\,T sweeps) or 2\,mK (3\,T, 4\,T sweeps, which were slightly more noisy) and calculated the mean value for each bin, thus obtaining resistivity at regularly sampled temperatures. Next, we applied a second order Savitzky-Golay filter, which works by fitting a polynomial of the corresponding order to a moving window. The second derivative is given by the curvature of this smoothing polynomial (a parabola in our case). The width of the moving window was either 50\,mK or 60\,mK (the larger window was used for the 3\,T and 4\,T sweeps). We then visually estimated the temperature interval which was guaranteed to contain the peak of the second derivative. The center of that interval was taken as \To, while its width defined the uncertainty.

\section{Expanded $T$-$H$ phase diagram}
\begin{figure}[ht!]
\includegraphics[width=\linewidth]{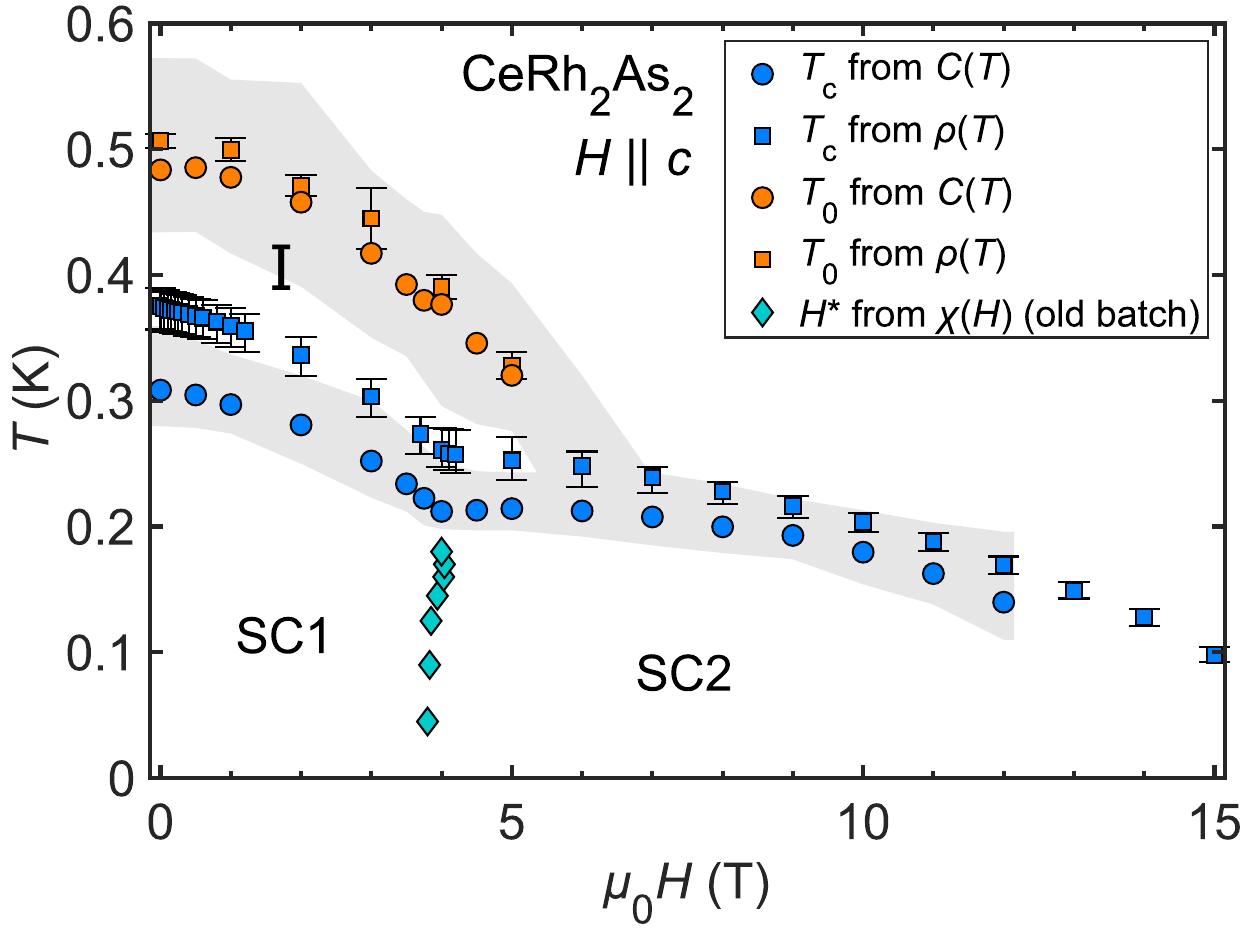}
\caption{\label{fig:SI_phase_diagram_expanded} The expanded version of the temperature - $c$-axis magnetic field ($T$-$H$) phase diagram of \CRA\ shown in Fig.~5 of the main paper. Definitions of the additional quantities and error bars plotted are explained in the text.}
\end{figure}

In Fig.~\ref{fig:SI_phase_diagram_expanded} we show the expanded version of the $T$-$H$ phase diagram of \CRA\ depicted in Fig.~5 of the main paper. In this version we added the \Tc\ values obtained from resistivity measurements. The associated error bars represent the width of the SC transition, defined as the interval between the 5\% and 95\% resistivity drop thresholds. The light grey area marks the onset-to-termination intervals for the \Tc\ and \To\ transitions in specific heat.
\section{Ginzburg-Landau theory of coupled order parameters}
The interplay between the superconductivity (SC) and the normal state ordering (phase-I) in \CRA\ near the tetracritical point of the $T-H$ phase diagram (Fig.~5 of the main article) can be analyzed in terms of a Ginzburg-Landau (GL) theory of coupled order parameters~\cite{imry1975,fernandes2010}. Considering spatially uniform order parameters for the two phases, $\Delta$ (SC) and $\mathbf{Q}$ (Phase I), the free energy is given as:
\begin{equation}
    F(\Delta,\mathbf{Q}) = \frac{a_{1}}{2}\Delta^{2} + \frac{b_{1}}{4}\Delta^{4} + \frac{a_{2}}{2}\mathbf{Q}^{2} + \frac{b_{2}}{4}\mathbf{Q}^{4} + \lambda\Delta^{2}\mathbf{Q}^{2}
\end{equation}
where $\lambda\Delta^{2}\mathbf{Q}^{2}$ is the coupling term. For $\lambda > 0$ the interaction is repulsive and represents competition between the two order parameters, while for $\lambda < 0$ the interaction is attractive resulting in supporting coupling (one enhances the other). 

We assume $T$-independent $b_{1},b_{2} > 0$, and close to the transition temperatures $T_{0}$ and $T_{c}$, we expand
\begin{align*}
    & a_{1}(T) = a_{1}(0)(T/T_{c} - 1)\\
    & a_{2}(T) = a_{2}(0)(T/T_{0} - 1).
\end{align*}

There are four solutions in zero field ($H = 0$): the non-ordered state (i), the SC state (ii), the phase-I state (iii) and the mixed state (iv):
\begin{align*}
i) ~ & ~ (\Delta_{0},Q_{0}) = (0,0)\\
ii) ~ & ~ (\Delta_{0},Q_{0}) = (\pm\sqrt{-a_{1}/b_{1}},0)\\
iii) ~ & ~ (\Delta_{0},Q_{0}) = (0,\pm\sqrt{-a_{2}/b_{2}})\\
iv) ~ & ~ (\Delta_{0},Q_{0}) = \left(\pm\sqrt{\frac{a_{1}b_{2}-2a_{2}\lambda}{4\lambda^{2}-b_{1}b_{2}}},\pm\sqrt{\frac{a_{2}b_{1}-2a_{1}\lambda}{4\lambda^{2}-b_{1}b_{2}}}\right)
\end{align*}
with condensation energies at $H = 0$:
\begin{align*}
i) ~ & ~ \Delta F = 0\\
ii) ~ & ~ \Delta F = -\frac{a_{1}^{2}}{4b_{1}}\\
iii) ~ & ~ \Delta F = -\frac{a_{2}^{2}}{4b_{2}}\\
iv) ~ & ~ \Delta F = -\frac{a_{1}^{2}b_{2}-4a_{1}a_{2}\lambda+a_{2}^{2}b_{1}}{4(b_{1}b_{2}-4\lambda^{2})}.
\end{align*}

We can now use the experimental values of the transition temperatures ($T_{c}$ ; $T_{0}$) and the jumps in the specific heat coefficients at the transitions ($A_{c}=\Delta C(T_{c})/T_{c}$ ; $A_{0}=\Delta C(T_{0})/T_{0}$) from the non-ordered state to the ordered states to calculate the parameters for every state independently. For instance, the specific heat change associated to the superconducting transition is
\begin{equation}
\Delta C(T_{c}) = \frac{a_{1}^{2}(0)}{2b_{1}T_{c}}.
\end{equation}
\begin{figure}[t]
\includegraphics[width=\linewidth]{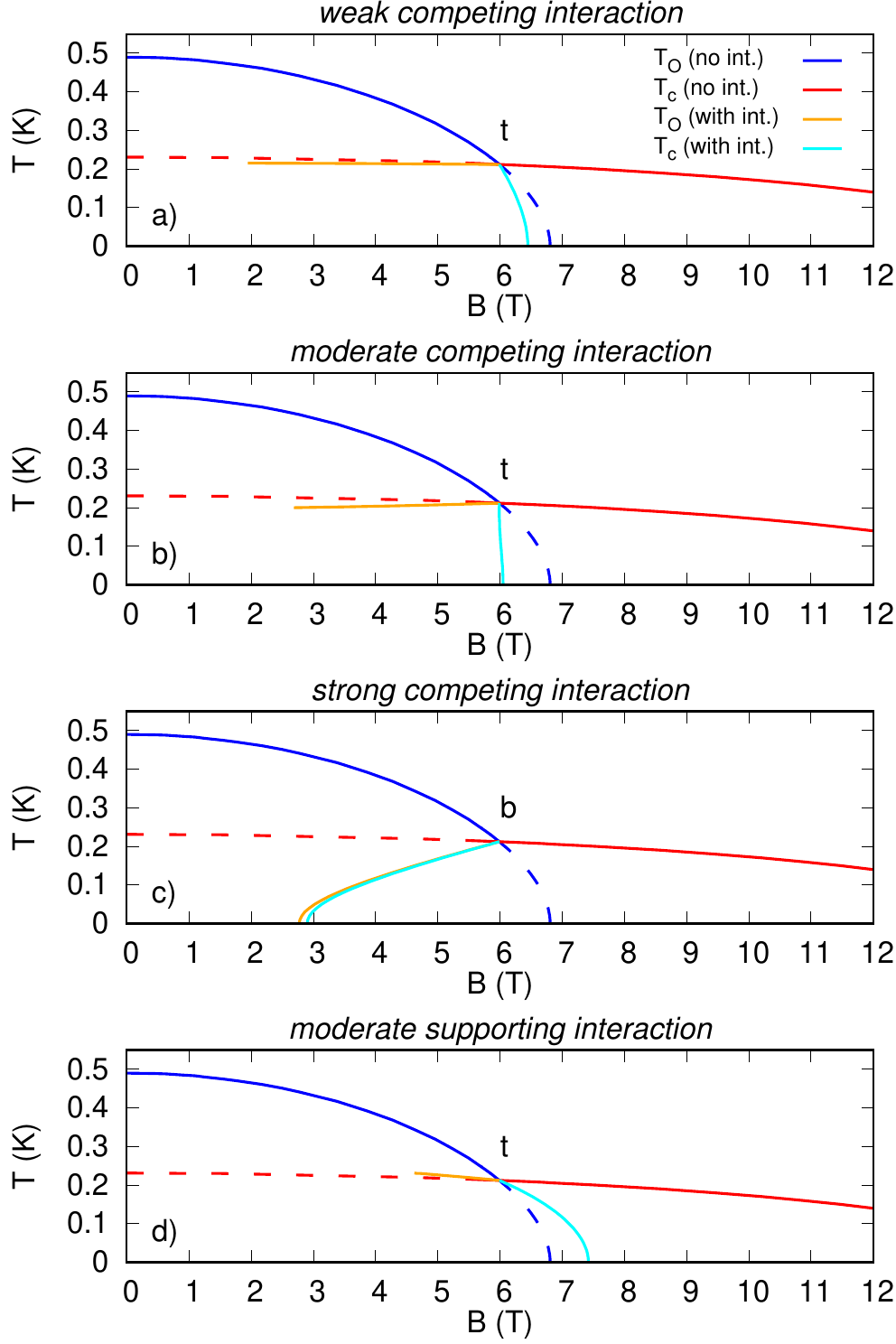}
\caption{\label{fig:SI_GL_theory} Temperature-field ($T$-$B$) phase diagrams for different values of the coupling paramater $\lambda$. $B = \mu_{0}H$ in tesla. The value of $\lambda$ for the weak coupling in panel a) is about 20\% of the value necessary to induce a first order phase transition as shown in panel c). The letter '\textit{t}' indicates the tetracritical point, whereas the letter '\textit{b}' indicates the bicritical point. The red and blue lines represent the phase boundaries without interaction (int.), with parts that get modified by the interaction drawn as dashed lines.}
\end{figure}

Assuming a power-law behavior for the specific heat of the form $C(T)/T = -\Delta\gamma + cT^{n}$ and inserting the magnetic field dependence as terms in the free energy as $a_{1}(0)\left(\frac{H}{H_{c}(0)}\right)^{m}\Delta^{2}$ and $a_{2}(0)\left(\frac{H}{H_{0}(0)}\right)^{m}\mathbf{Q}^{2}$, it is possible to reproduce the experimental phase diagrams for the SC2 and phase-I states of \CRA\ without interaction ($\lambda = 0$) and extract the $n$ and $m$ parameters. These are plotted in Fig.~\ref{fig:SI_GL_theory} for $T_{c} = 0.22$\,K and $T_{0} = 0.49$\,K as red and blue lines, respectively.

By solving the GL equations with the coupling term and considering that at the '\textit{t}' point ($H_{t},T_{t}$) of the phase diagram in Fig.~5 of the main article, $T_{c}(H_{t}) = T_{0}(H_{t})$, one can draw the phase diagrams for different values of the coupling paramater $\lambda$. In fact, performing a similar calculation as in Ref.~\cite{fernandes2010}, one derives simple expressions for the field dependences of the transition temperatures:
\begin{align}\label{eq:boundaries}
    & T_{c}(H) = T_{t}\left[ 1 + \frac{f(H)}{1-\frac{\Gamma}{2A_{c}(T_{t})}}\right] ~ \mathrm{for} ~ H < H_{t}\\
    & T_{0}(H) = T_{t}\left[ 1 + \frac{f(H)}{1-\frac{2A_{0}(T_{t})}{\Gamma}}\right] ~ \mathrm{for} ~ H > H_{t}
\end{align}

where we assume $T_{0}(H) = T_{t}[1 + f(H)]$ for $H < H_{t}$, $df(H)/dH < 0$, $f(H_{t}) = 0$ and $T_{c}$ independent of $H$ for $H > H_{t}$. $\Gamma$ is a scaled strength of the interaction whose energy is $E_{int.} = \Gamma T_{t}^{2}$ similar to the condensation energies of both phases being $E_{c} = A_{c}T_{t}^{2}$ and $E_{0} = A_{0}T_{t}^{2}$. The phase boundary lines from these equations are plotted in Fig.~\ref{fig:SI_GL_theory} as orange and cyan lines.

\begin{figure}[b]
\includegraphics[width=\linewidth]{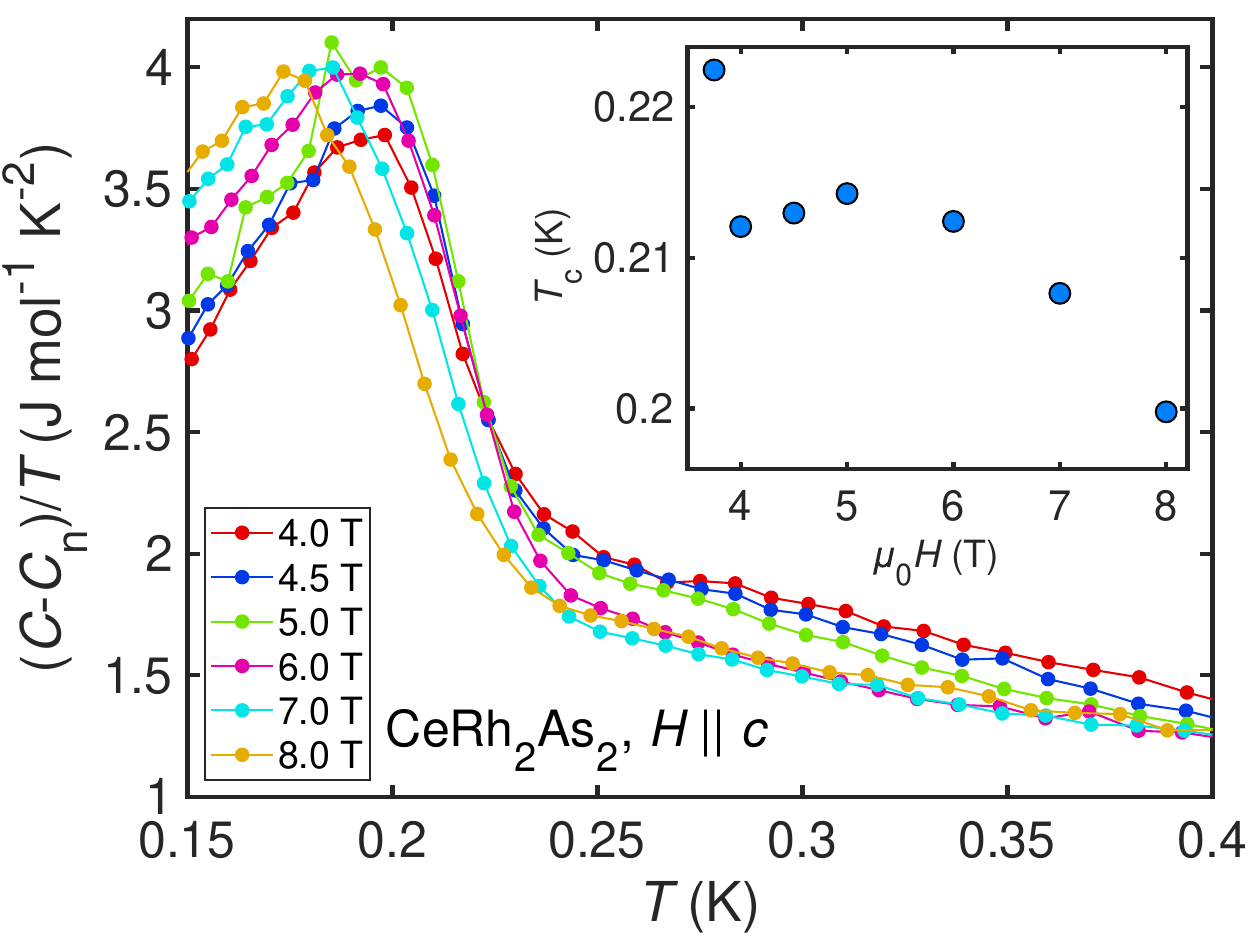}
\caption{\label{fig:SI_Tc_slope_change}Main plot: temperature ($T$) dependence of electronic specific heat $(C-C_{\mathrm{n}})/T$ of \CRA\ at specific values of the $c$-axis magnetic field ($H$) between 4\,T and 8\,T. Inset: field dependence of the superconducting transition temperature $T_{c}$ in a similar field region.}
\end{figure}

We now describe the results starting from the panel a), i.e., weak coupling. By weak coupling we mean a value of $\lambda$ which is 20\% of that necessary to induce a first order phase transition (case shown in panel c)). We have four second order phase transition lines joining at $t=(H_{t},T_{t})$, the tetracritical point. Although the coupling is weak a substantial change in slope of the $T_{0}(H)$ line is observed below the tetracritical point (blue to cyan line). The line becomes steeper, and with a moderate coupling (50\% relative to the case in panel c)), it becomes almost vertical (shown in panel b)). This could explain why we do not observe this line in the experiments, for instance we do not detect any transition in specific heat at 7\,T. In contrast, the phase line of the superconducting phase for $H < H_{t}$ is not affected at all. This is due to the fact that the slope is proportional to the specific heat jumps $A_{c}(T_{t})$ and $A_{0}(T_{t})$ (cf. equations above) evaluated at the tetracritical point and experimentaly, $\Delta C(T_{0})/T_{0} \approx \frac{1}{10}\Delta C(T_{c})/T_{c}$.

A strong coupling is expected to induce a first order transition between the SC and phase-I states as shown by the orange/cyan line in panel c) of Fig.~\ref{fig:SI_GL_theory}. The joining point is therefore a bicritical point '\textit{b}'. This is definitely not the case in \CRA.

Finally, a supporting interaction ($\lambda < 0$) results in an increase of $T_{0}(H)$ towards higher fields (shown in panel d)), in which case the \To\ transition should have been observed in our experiments at 7\,T.

A careful look at our experimental data reveals what seems to be a small change in the slope of the $T_{c}(H)$ line at about 6\,T. As emphasized in Fig.~\ref{fig:SI_Tc_slope_change}, \Tc\ remains more or less constant in the 4\,T to 6\,T range and clearly decreases with field for $H>6$\,T. While resolution of our data is insufficient for a reliable determination of the slope change, the observed behavior decently matches the one predicted for the weak competing coupling regime. We therefore propose that this particular case describes the interaction between the SC and phase-I orders in \CRA. 
\bibliography{semeniuk_prl_20230403}
\bibliographystyle{apsrev}
\end{document}